# Online handwriting, signature and touch dynamics: tasks and potential applications in the field of security and health

**Marcos Faundez-Zanuy · Jiri Mekyska · Donato Impedovo**

**Abstract** Background: An advantageous property of behavioural signals (e.g. handwriting), in contrast to morphological ones, such as iris, fingerprint, hand geometry, etc., is the possibility to ask a user for a very rich amount of different tasks.

Methods: This article summarises recent findings and applications of different handwriting/drawing tasks in the field of security and health. More specifically, it is focused on on-line handwriting and hand-based interaction, i.e. signals that utilise a digitizing device (specific devoted or general-purpose tablet/smartphone) during the realization of the tasks. Such devices permit the acquisition of on-surface dynamics as well as in-air movements in time, thus providing complex and richer information when compared to the conventional "pen and paper" method.

Conclusions: Although the scientific literature reports a wide range of tasks and applications, in this paper, we summarize only those providing competitive results (e.g. in terms of discrimination power) and having a significant impact in the field.

## 1 Introduction

Signature/handwriting recognition can be split into two categories: off-line and on-line [2, 24, 90]. In the former case, just the result of the signature/writing (i.e. static

M. Faundez-Zanuy
Escola Superior Politecnica, Tecnocampus, Avda. Ernest Lluch 32, 08302 Mataro, Barcelona, Spain
E-mail: faundez@tecnocampus.cat

J. Mekyska
Department of Telecommunications, Brno University of Technology, Technicka 10, 61600 Brno, Czech Republic

D. Impedovo
Dipartimento di Informatica, Università degli Studi di Bari Aldo Moro, Via E. Orabona 4, 70125 Bari, Italy

2D image) is known because it is acquired after the realization (writing) process. On the other hand, online signature/writing consists of acquiring the signal during the realization process. This provides a large set of raw data:

- absolute spatial coordinates (x, y) of the tip of the pen,
- pressure exerted on the surface – of course, this value is zero when the pen is not touching the surface,
- angles of the pen: altitude and azimuth,
- time stamp of the moment where the previous values have been acquired.

When pressure is different from zero the movement is considered to be on-surface and the whole set of information described before is acquired. When pressure is zero, the movement is considered to be in-air. If the distance from the tip of the pen to the paper surface is below one centimeter (depending by the specific acquisition tool) the whole set of information described before is acquired with the unique exception of pressure, which is always zero. A deeper discussion linked with the in-air movement can be found in our previous works [90, 2, 3].

From a pattern recognition perspective, off-line systems deal with image processing, while on-line ones with time-sequence signal processing. However, it must be argued that, so far, some solutions developed for off-line systems have been adopted to on-line one and vice-versa [46]. An emerging and very interesting aspect dis-cussed in this article deals with the possibility to sign and/or write and more in general interact with the finger on a screen of a mobile device (e.g. smartphone or tablet) [47].

Four components are embedded in the signing/writing/drawing process [47, 48]:

- The physiologic component is constituted by the writing system which includes muscles, arm, wrist, hand, fingers, etc.;
- The learned component deals with personalization, schooling, culture, habits, etc.;
- The cognitive one can be referred to mental abilities (learning, thinking, reasoning, remembering, problem-solving, decision-making, and attention);
- The contour component: given the above, some noise can be introduced due to the writing device, posture, spatial constrains, emotional state, etc.

Variations of these components are reflected into variations of the acquired signal and represent the intra-writer variability. The variability is then, typically, observed over short (day-to-day or trial-to trial basis) or long periods (months, years, etc.). In the former case, the contour component has a major role on the overall variability [47], while in the latter one all the different components could have a significance and different impact [48]. It is quite intuitive that the handwriting signal can be used for multiple purposes: handwriting recognition [89], script recognition, drawing recognition [62], health evaluation, assessment of specific learning disabilities, gender recognition [91], fatigue detection [34], emotional state recognition [57], forensic studies, writer identification (based on signature or handwriting) and signature/writer verification.

However, handwriting does not only include the writing of cursive/capital letters or scripts, in fact drawings can be considered too. More specifically, the different handwriting tasks can be classified as [48]:

1. Simple drawing tasks: straight lines, circles, spirals, meanders, swipes, etc. These tasks are also referred to as graphomotor elements, because they represent the basic building blocks of letters;
2. Simple writing tasks: nonsense words, single characters, single tap, etc.;
3. Complex tasks: they simultaneously involve motor, cognitive, and functional issues (e.g. copying tasks, the clock drawing task, etc.).

It has been demonstrated so far that, given a specific classification problem (e.g. writer identification, health status assessment, etc.), a specific task is more profitable than others. In fact, intuitively, given a specific writing task, one of the previous mentioned components could have a different impact on the acquired signal.

Automatic handwriting-based personal identification can be based in three different tasks, which will be described in detail in the next subsections.

## 2 Handwritten tasks

### 2.1 Signature-based analysis

Figure 1 shows an example of a signature acquired with a Wacom Intuos digitizing tablet. The blue color represents the on-surface movement, while the red color the in-air one. The relevance of the in-air movement has been clearly described in [24, 90, 49]. Handwritten signature is the most widespread behavioral biometric traits: it has a social accepted role as proof of identity as well as a demonstration of the willing of the writer in accepting the content of the document. For this reason, it has been extensively analyzed [108]. Signature is adopted on a day basis for commercial and banking payments/transactions and in many other sectors (e.g. express courier, education, healthcare, etc.). Several international competitions exist that facilitate a fair comparison between competing algorithms [89].

Although it is not massively used in health applications sometime interactions appear between security and health, such as in documents signed by a user suffering dementia or some other temporary/permanent health problem that can invalidate the signature. An example has been reported in [25]. The interesting aspect is that usually, security and health implications are present both together and cannot be considered isolated application fields [25].

Micrographia (the abnormal progressive reduction in amplitude of letters) has been observed in the off-line and on-line signing tasks as well as in sentences of patients with Parkinson's disease [104, 113]. Signature position with respect to a dotted line (on or below) and other cognitive functions have been investigated, and it has been observed that it may be a marker of vulnerability for visuospatial abilities [109].

Recently, it has been demonstrated that, when dealing with on-line writing, velocity-related features play a very crucial role [45, 71]. A similar result has been observed on signatures when considering features related to the Sigma-Lognormal model coupled with a Bagging CART classifier [79]. In this case, the approach has been able to discriminate dementia affected users from healthy counterpart with 3% of Equal Error Rate (ERR), however the main limitation of this work relies in a reduced dataset. A more recent study has investigated the relation

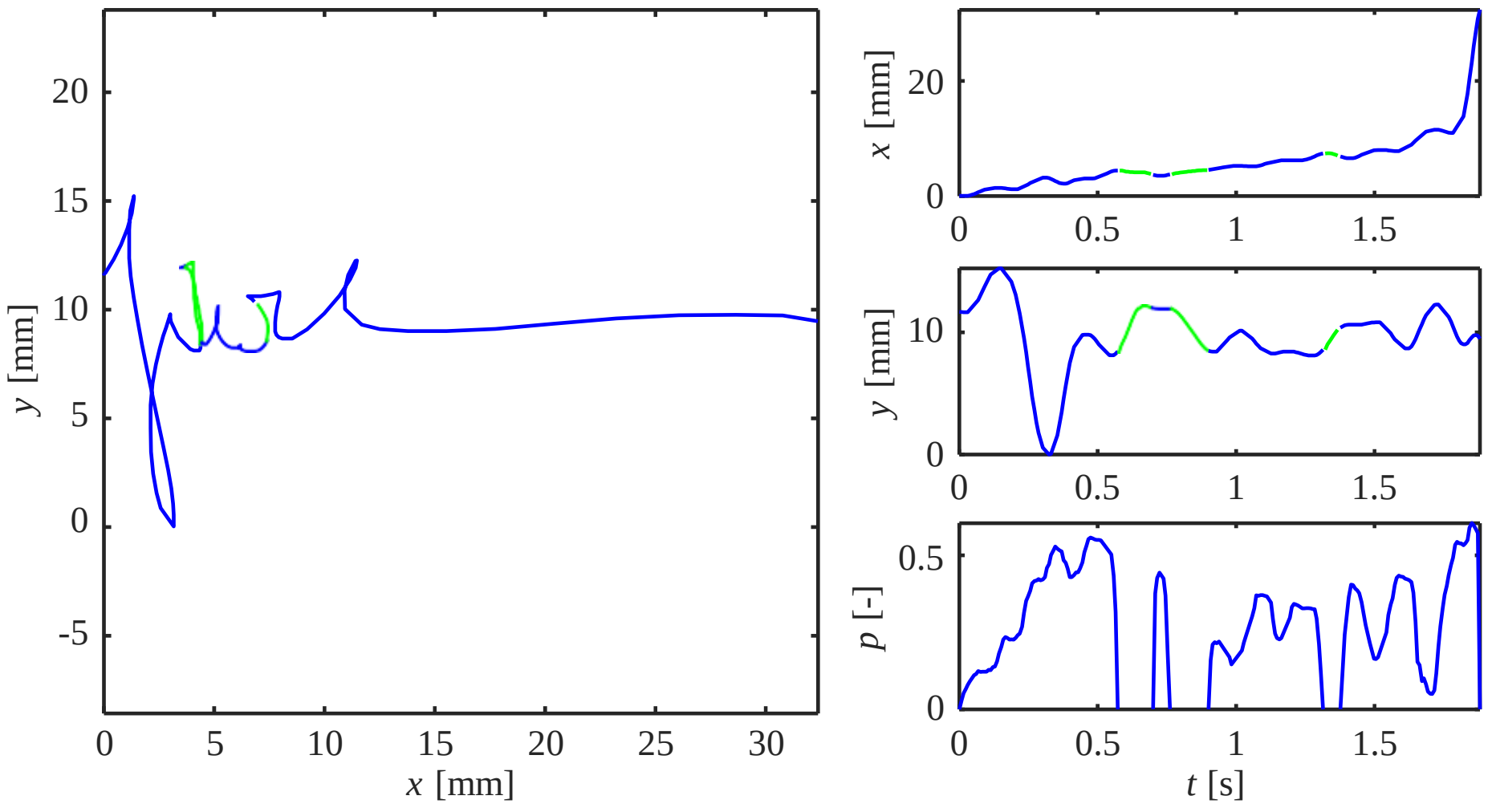

Fig. 1: Online signature: the product is depicted on the left side, the right side of the figure contains associated information about horizontal/vertical movement and pressure pattern (the blue color represents the on-surface movement, while the green color the in-air one)

between signatures of persons with Alzheimer's Disease (AD) and those written by age-matched Healthy Controls (HC) [8]. In this case authors adopted simple statistical evaluations on parameter features evaluated upon dynamic raw data (e.g. stroke duration, stroke amplitude, peak vertical velocity, average normalized jerk, etc.) and categorized signatures within three classes: text-based, mixed, or stylized. No significative differences were observed apart from an association between increased feature variability and increased dementia severity for stylized and mixed signatures. Signatures were also acquired after one year during which a hard decline was observed in the cognitive status: signature features remained stable. Authors conclude the work stating that dementia has residual impact on signature formation. A similar finding is reported by Reiner et al. [82] which acquired two samples of signature and a spontaneous writing from 36 persons with Mild Cognitive Impairment (MCI) diagnosis and 38 HC. Cognitive functions in decision-making were also evaluated: while a significant correlation between spontaneous writing and neuropsychological test results was observed, signature deterioration did not appear to be correlated with the level of cognitive decline. However, it must be underlined that the style of the signature plays a role, in fact the speed for flourish signatures is higher than that of text-based ones, moreover muscles involved in the movement are more active in the generation of the flourish [10]. These results call for further and extended research.

The relation between handwritten signatures and personality traits has been evaluated considering static and dynamic features. It is interesting to report that aspects as gender and can be predicted effectively using signature velocity characteristics [68].

On-line signatures have been also used (coupled with speech) to distinguish among three psychophysiological states: normal, drowsiness and alcohol-intoxicated [64].

Dynamic and static features were adopted to test Bayesian hypothesis reporting an overall average error of 14.5%.

Unfortunately, very few works provide a comparison of performance when adopting writing, drawing and signing tasks. From an intuitive point of view, handwriting should be able to provide a wider range of information. More evidences are reported in the following sections.

## 2.2 Text-based recognition analysis

An example of a cursive handwriting could be seen in Figure 2. Several security applications based on handwritten text have been proposed, such as [89] for capital letters or [88] for cursive drawing in a whiteboard, which is not a very usual writing scenario. However, they have not attracted too much the attention of the scientific community. Especially when compared to signatures.

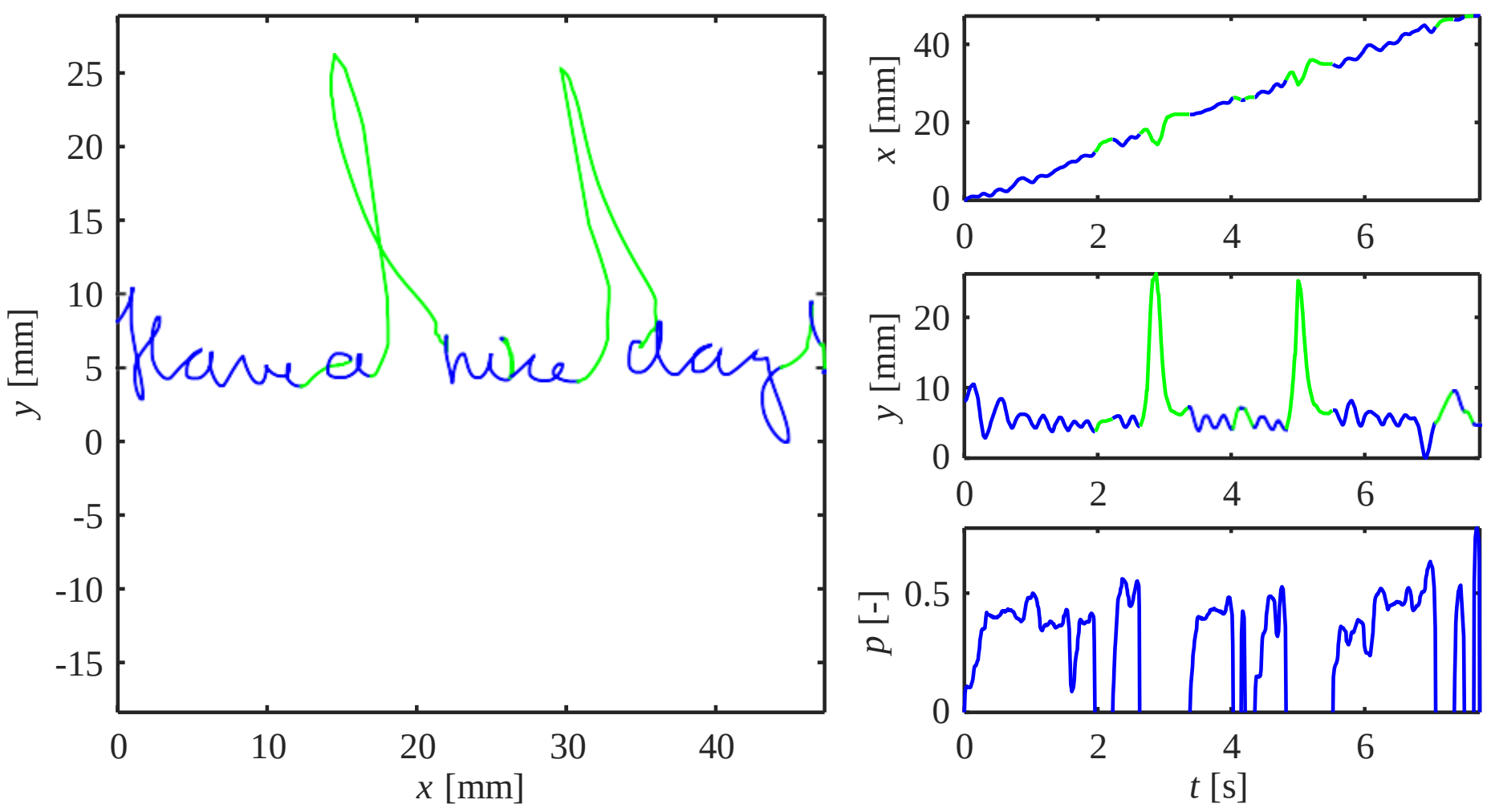

Fig. 2: Text "Have a nice day!" written in cursive letters: the product is depicted on the left side, the right side of the figure contains associated information about horizontal/vertical movement and pressure pattern (the blue color represents the on-surface movement, while the green color the in-air one)

## 2.3 Drawing-based analysis

In security applications it has attracted some attention specially in graffiti performed by gangs. Gangs use specific clothing, brands, symbols, tattoos, and graffiti to identify their group and interchange messages. Graffiti are any type of public markings that may appear in forms that range from simple written words to elaborate wall paintings [51]. However, due to its nature they are off-line. Preliminary results in online recognition show a potential to identify people using some drawings [62].

Generally, in the area of diagnostics in medical context, drawings are widely used. Some common drawings and their potential usage in medical field are mentioned below.

### *2.3.1 Pentagon test*

The test is used e.g. in the Mini Mental State Examination (MMSE) to assess cognitive impairment [30]. It consists of copying a drawing, which includes two pentagons overlapping into a rhombus (see 3). It is of interest to report that it has been adopted to differentiate dementia associated with Lewy Body (DLB) from Alzheimer's Disease (AD). To the aim, visual parameter features such as number of angles, distance/intersection, closure/opening, rotation and closing-in where considered with an Artificial Neural Network classifier [6]. Park et al. [75] have recently adopted a mobile device to acquire timestamps, x-y coordinates and touch-events. In this case, raw data were processed by means of U-Net (a convolutional network) to automatically segment angles, distance/intersection between the two drawn figures, closure/opening of the drawn figure contours. Moreover, tremor was also evaluated. It is worth noting that the evaluation of these parameters is associated with a specific disease scaling (interested readers can refer to [6]). Errors which occur in the copying/drawing tasks can be related to damages into the brain: it has been observed that the score connected to the pentagon copy task is associated with parietal grey matter volume and not with frontal, temporal, and occipital ones [94].

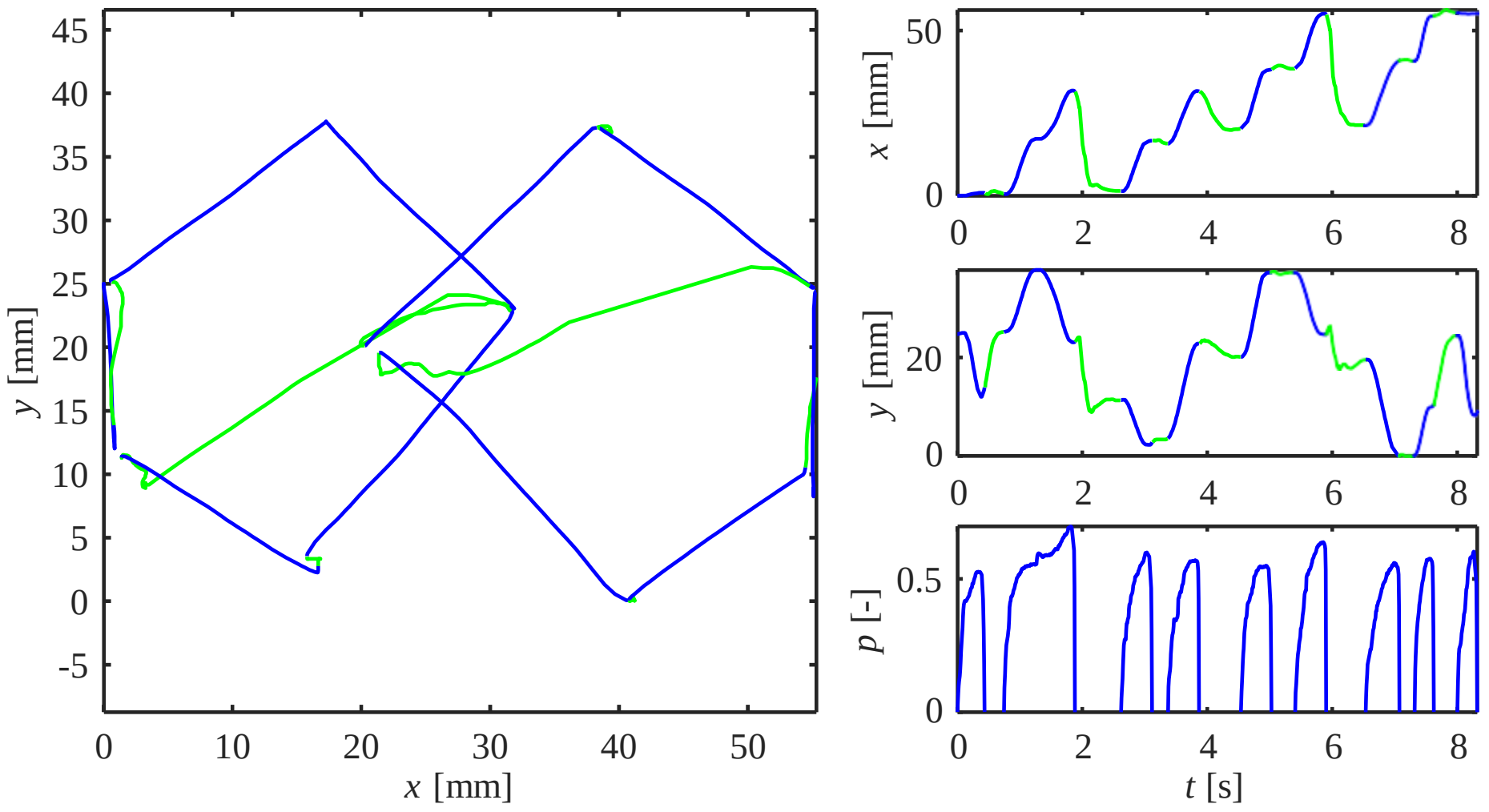

Fig. 3: Pentagon test: the product is depicted on the left side, the right side of the figure contains associated information about horizontal/vertical movement and pressure pattern (the blue color represents the on-surface movement, while the green color the in-air one)

### 2.3.2 *Clock drawing test (CDT)*

The test can be utilized as a precursory measure to indicate the likelihood of further/future cognitive deficits. It is used e.g. in the Addenbrooke's Cognitive Examination – Revised (ACE-R) test [55] (see Figure 4). The use of an on-line acquisition tool gives the possibility to evaluate the process of the clock construction and not only the final drawing. In the last decade digital on-line versions of the CDT have been considered [52]. Harbi et al. [40] used a set of features extracted at stroke level (evaluated upon the set of on-line raw data) and a SVM to identify connected components in normal and abnormal drawings. The same authors also proposed a multi-expert approach [39]. More specifically two systems were developed: the first one considered static images obtained by plotting the x-y coordinates and derived a set of static features evaluated by means of a CNN. The same CNN provided a final decision. The second system was based on the x-y coordinates sequences. It was showed that the combination of both systems was able to outperform individual classifiers in the dementia vs heathy subject classification task. Muller et al. [73] investigated the diagnostic value of a digital version of the CDT by comparing it to the standard pencil-paper version. To the aim, 20 patients with early dementia, 30 with MCI and 20 HC were considered. It was observed that in-air time provided by the digital version is able to provide a higher diagnostic accuracy (MCI vs HC) than the use of the traditional test.

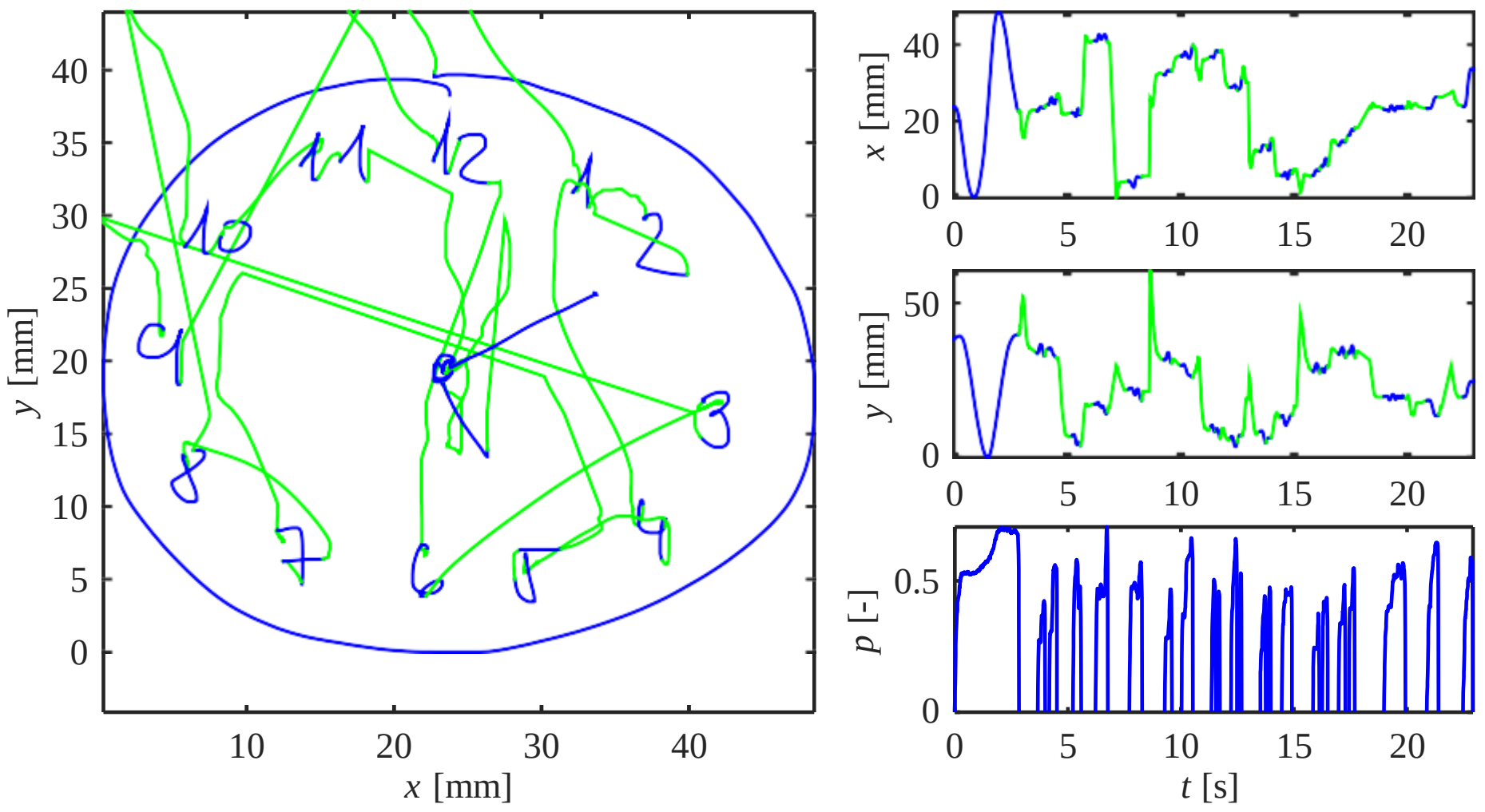

Fig. 4: Clock drawing test: the product is depicted on the left side, the right side of the figure contains associated information about horizontal/vertical movement and pressure pattern (the blue color represents the on-surface movement, while the green color the in-air one)

### 2.3.3 House drawing copy

This test is used for identification of Alzheimer's disease [35, 26] (see Figure 5).

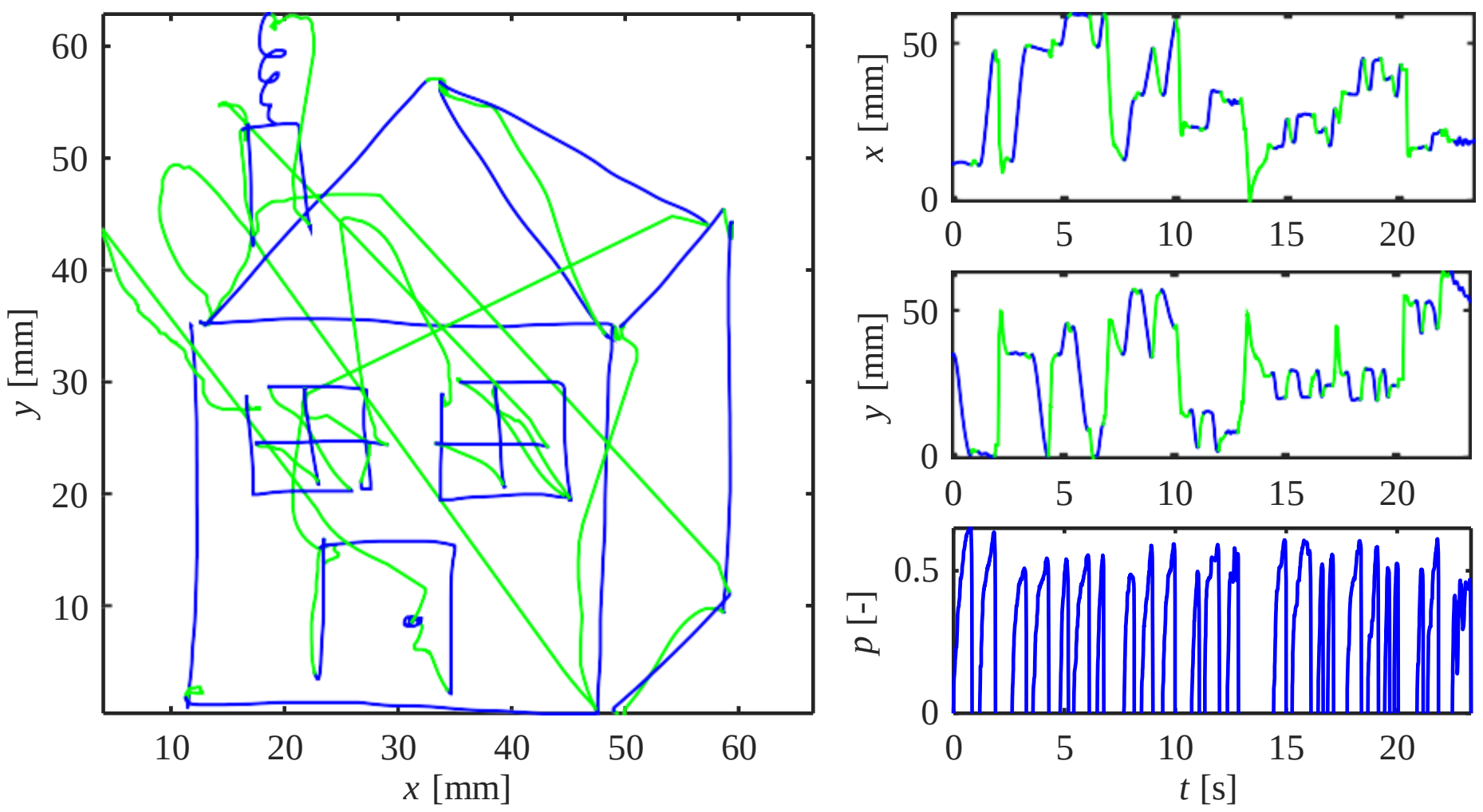

Fig. 5: House drawing test: the product is depicted on the left side, the right side of the figure contains associated information about horizontal/vertical movement and pressure pattern (the blue color represents the on-surface movement, while the green color the in-air one)

### 2.3.4 Archimedean spiral and straight line (drawing between points)

These tasks are useful to support diagnosis of e.g. Parkinson's disease [86, 85], Huntington's disease [7], essential tremor [63, 80, 70, 27], developmental dysgraphia [31], fatigue [34], or brachial dystonia [81]. See Figure 6. In the case of the Archimedean spiral acquisition and straight lines, the participants can have a printed spiral on a sheet of paper and a couple of dots to be connected and they are asked to trace it by a pen without touching the spiral neither the bars (see Figure 7). Or, the spiral is shown them on a template and they are asked to replicate it on a blank sheet of paper.

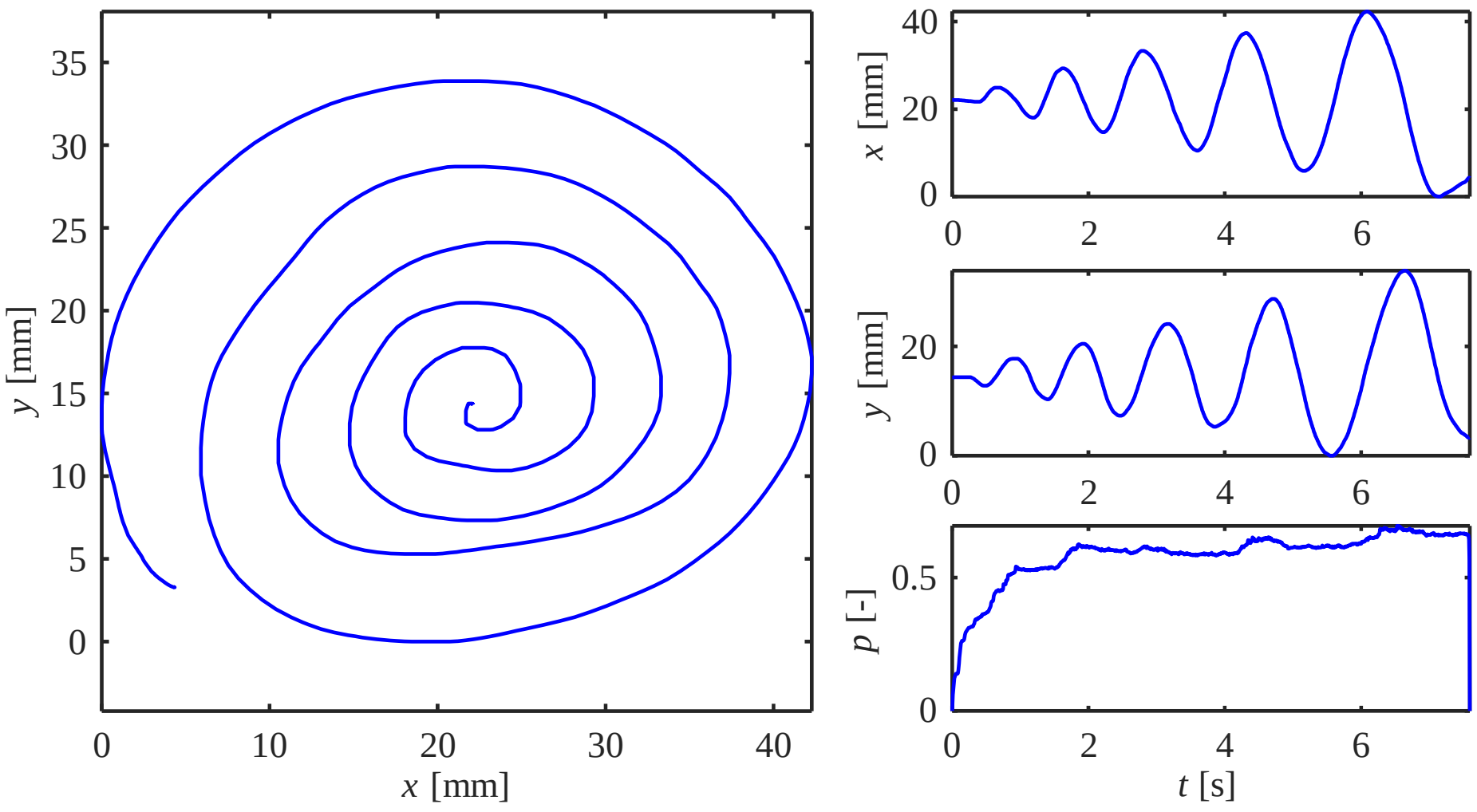

Fig. 6: Archimedean spiral: the product is depicted on the left side, the right side of the figure contains associated information about horizontal/vertical movement and pressure pattern

(a)

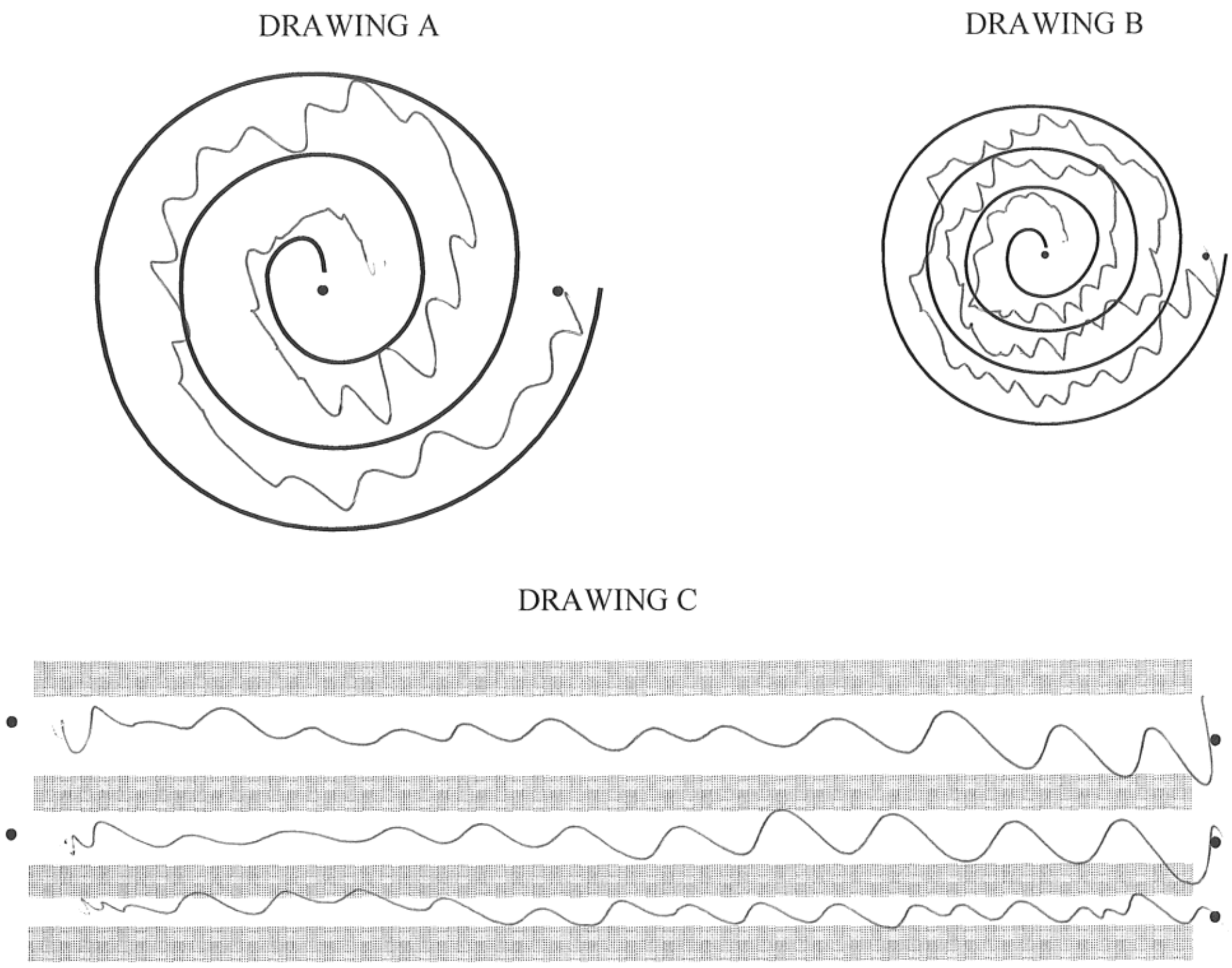

(b)

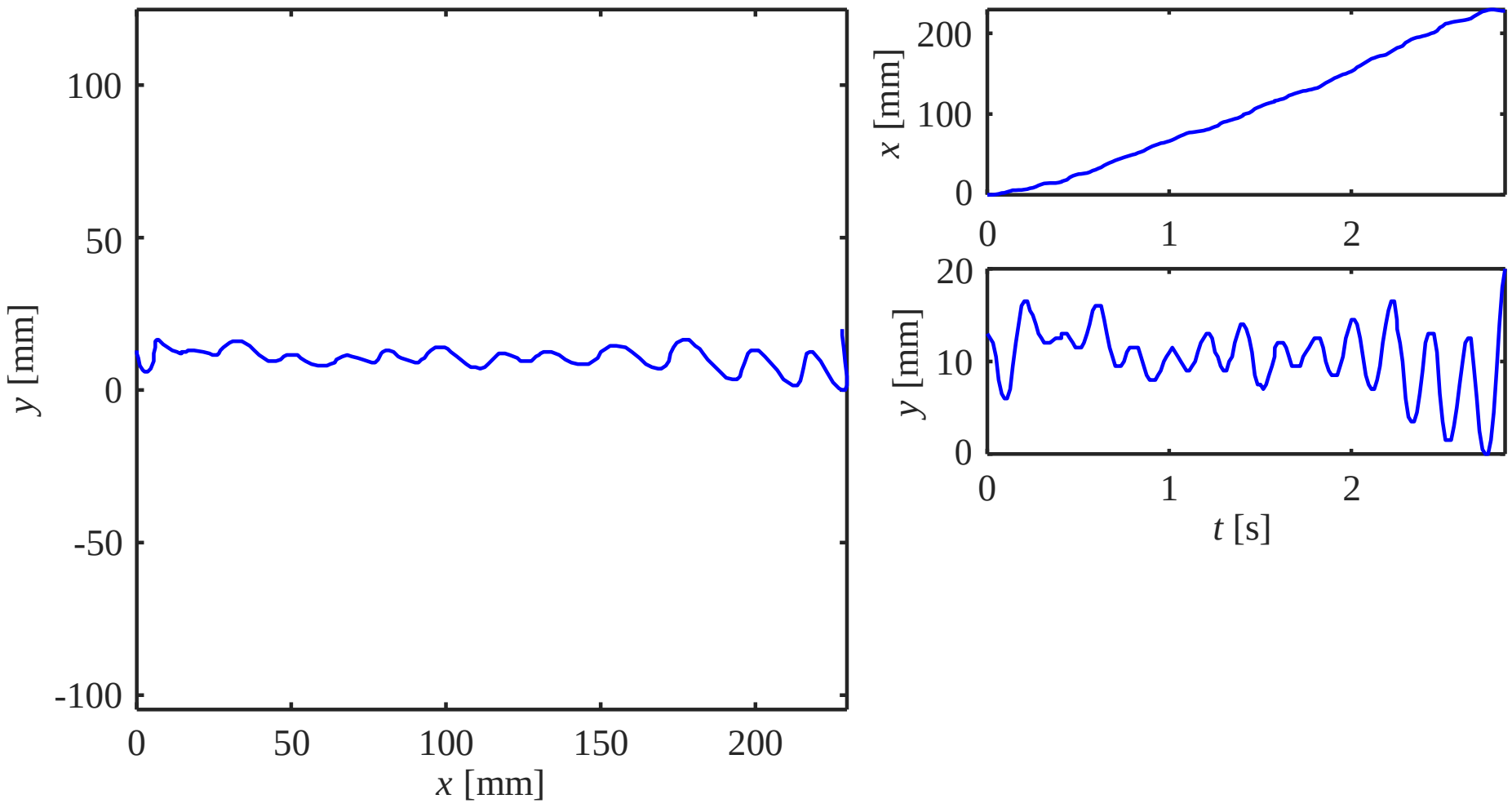

Fig. 7: a) Archimedean spirals and straight lines performed by a subject with essential tremor on a sheet of paper; b) reconstruction of the first straight line (information about the pressure is missing)

### 2.3.5 Overlapped circles (ellipses)

It can be used for quantitative analysis of schizophrenia or Parkinson's disease [9, 71]. See Figure 8, which represents some simple kinematic features that can be used for an effective diagnosis.

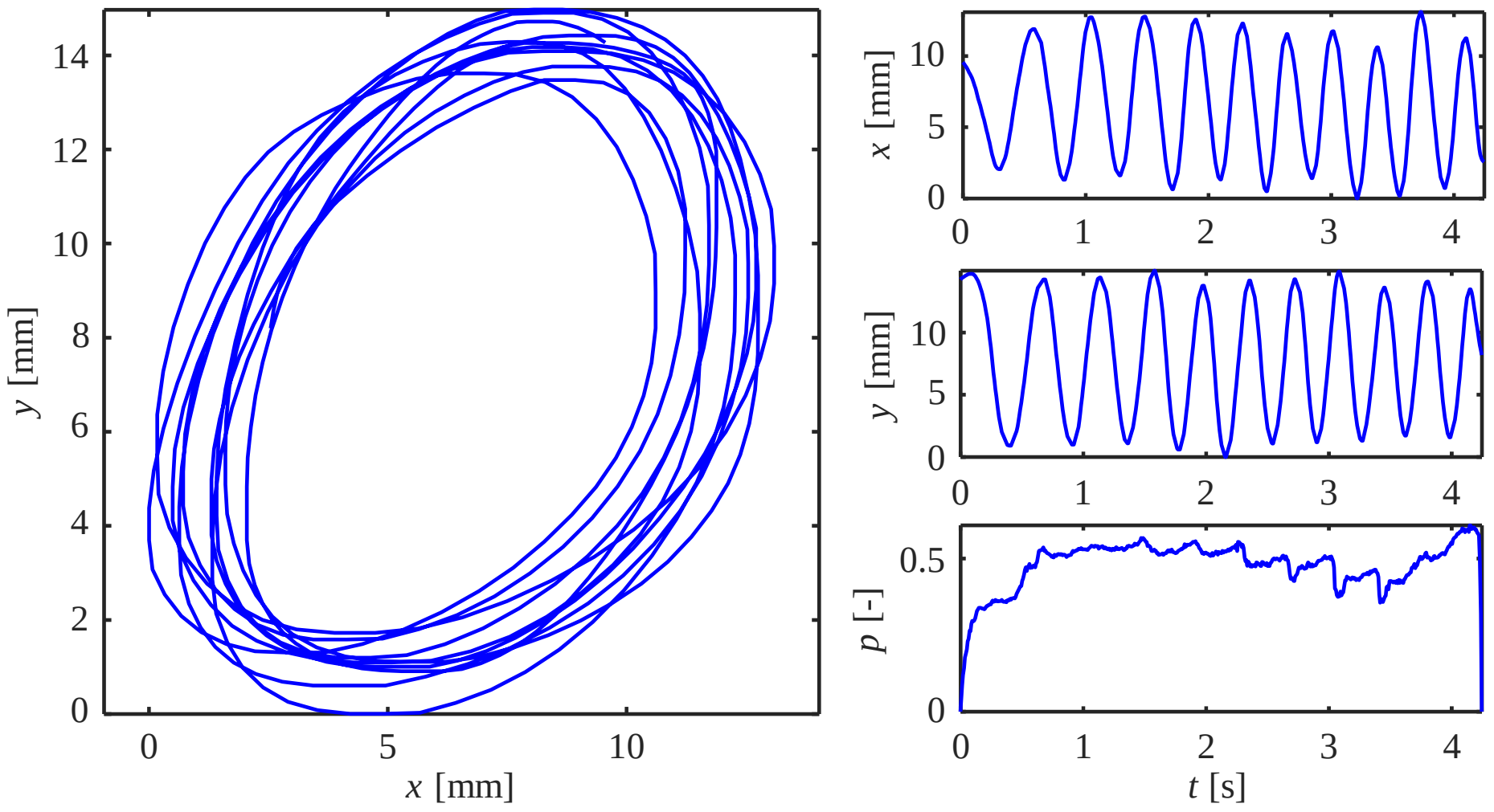

Fig. 8: Overlapped circles: the product is depicted on the left side, the right side of the figure contains associated information about horizontal/vertical movement and pressure pattern

### 2.3.6 Spring task (connected l or loops)

Several variants exist, such as the connected loops (see Figure 9), inverted connected loops, connected f, etc. This task is interesting to check the skill to produce rhythmic movements, as well as sudden changes of direction (start-stop-start sequences), useful to evaluate problems to initiate movement [93].

### 2.3.7 Rey–Osterrieth complex figure test (ROCF)

It was developed in 1941 and further consists of copying a complex drawing [13]. It is frequently used to further explain any secondary effect of brain injury in neurological patients, to test for the presence of dementia, or to study the degree of cognitive development in children. In this task patients have to memorize an image and later they have to replicate it without looking at the example. Rey–Osterrieth complex text is depicted in Figure 10.

### 2.3.8 Bank-check copying

It is, as for most cases of copying tasks, a functional writing task. To properly copy the bank check (Figure 11), the user should be able to identify the source and corresponding target fields, to locate them and to write the correct content. The single movement and the corresponding stroke could be correct, but the task must be evaluated in its total execution. Patients affected by dementia could result in poor execution producing simplified figures, miss-placement of the text, modifications in spatial relationships among strokes, etc. [16]. Considering the example reported in Figure 11, it can be observed clearly in-air movements which reveal the action of locating the source and the corresponding target field performed by a mild stage dementia patient.

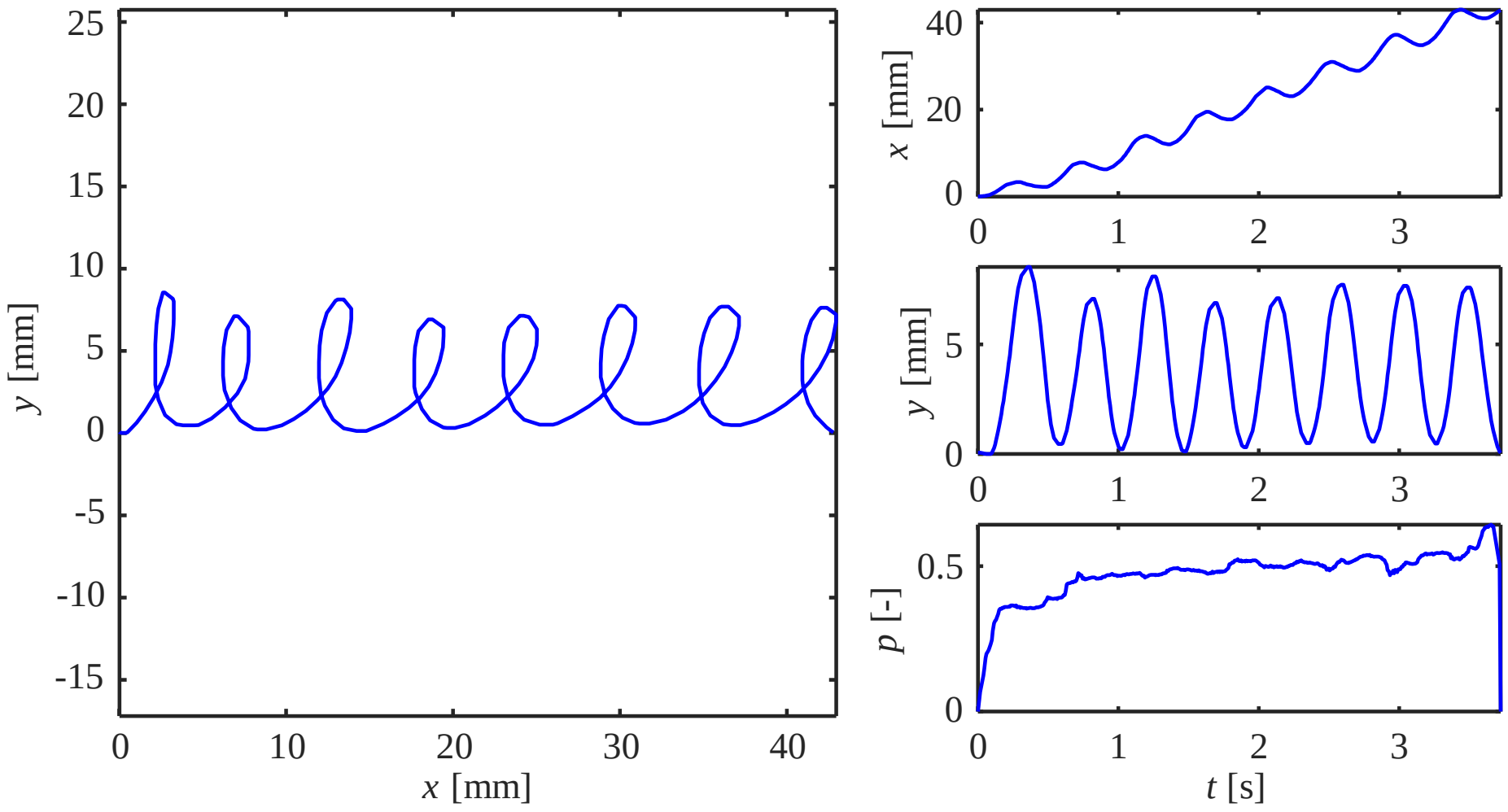

Fig. 9: Spring task: the product is depicted on the left side, the right side of the figure contains associated information about horizontal/vertical movement and pressure pattern

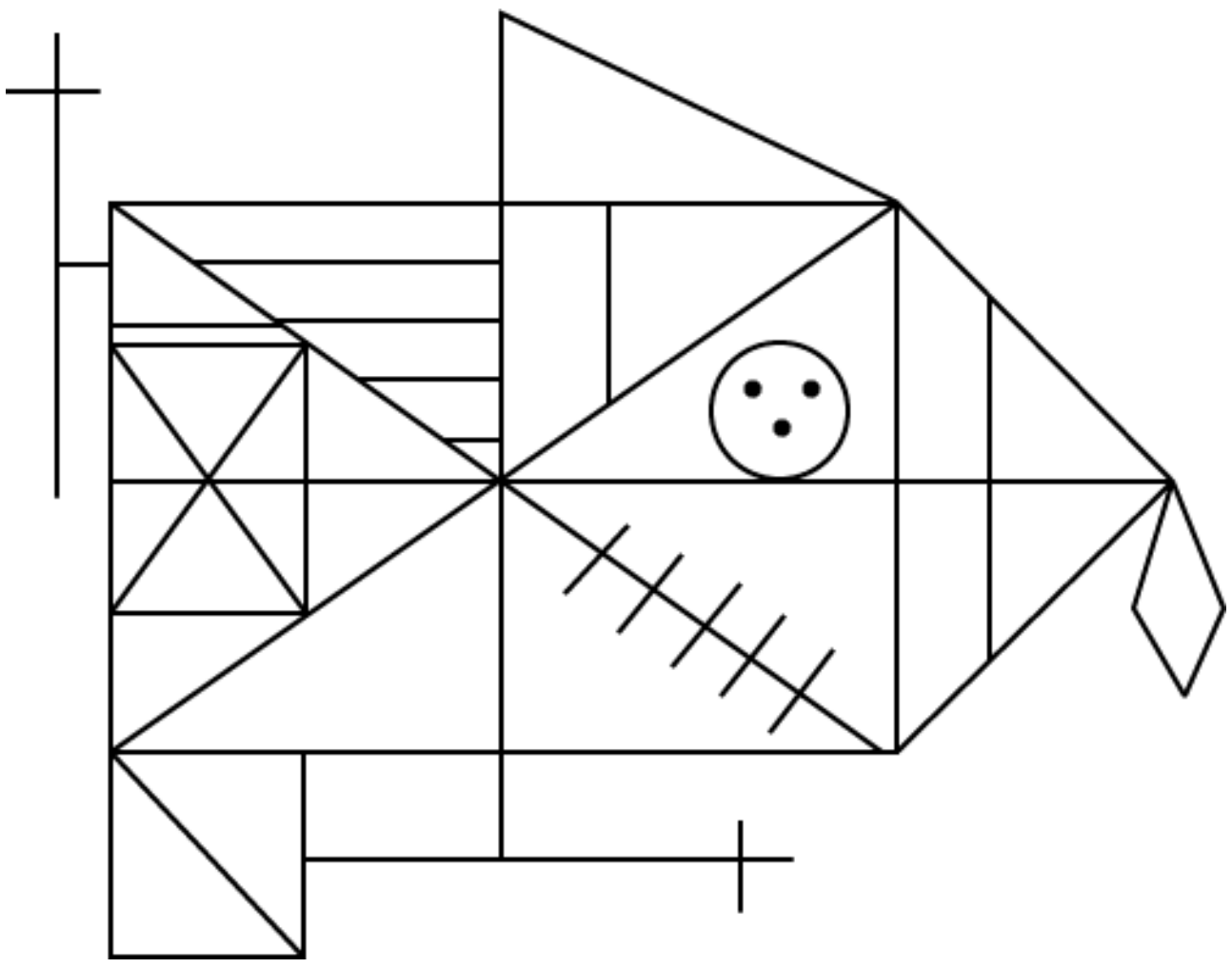

Fig. 10: Rey-Osterrieth complex figure test

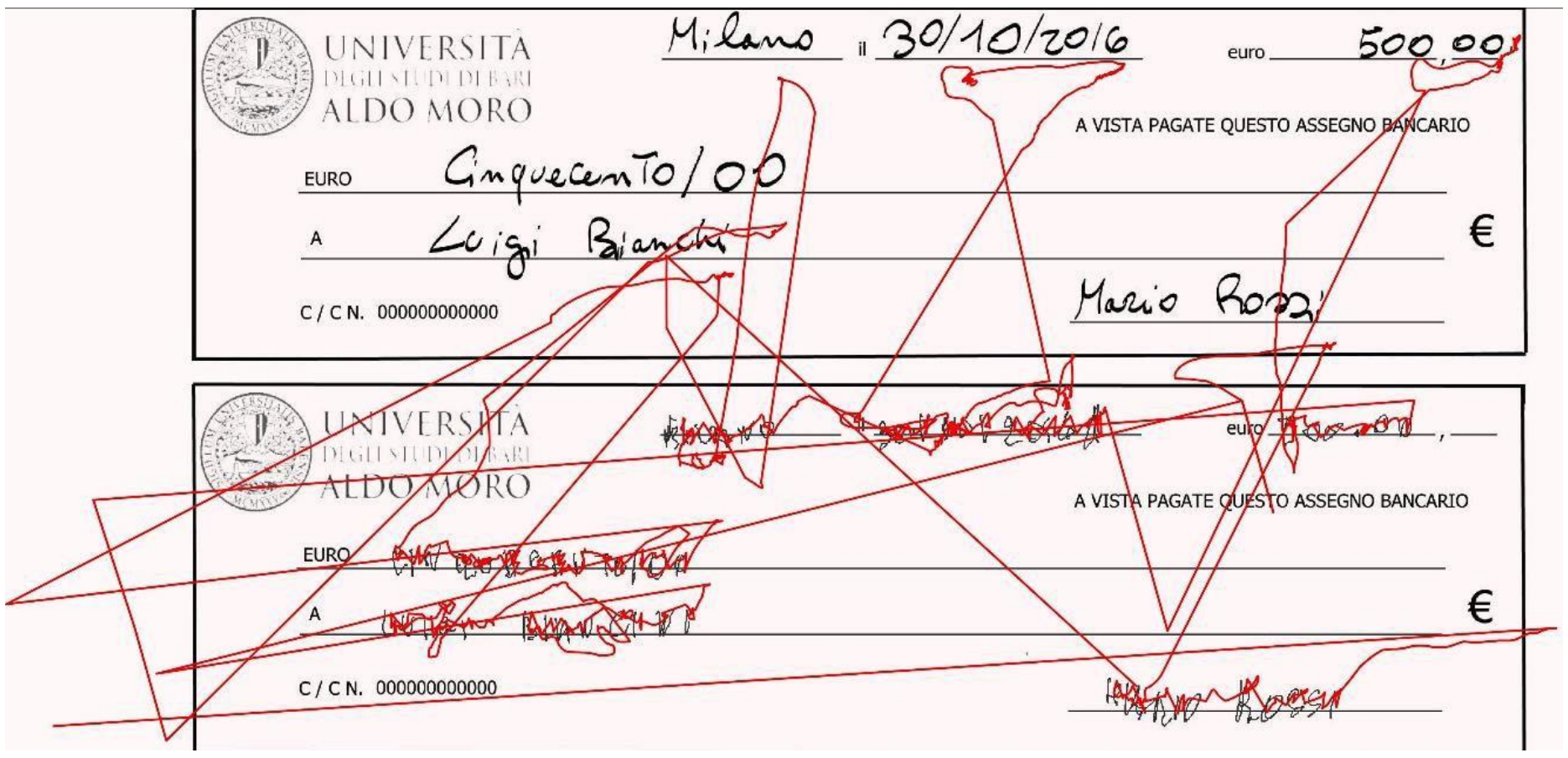

Fig. 11: Bank-check copying (the black color represents the on-surface movement, while the red color the in-air one)

### *2.3.9 Trail making test (TMT)*

The test is composed by two parts, in part A the user is required to connect a sequence of consecutive numeric targets (Figure 12), in part B numbers and letters must be alternated in progressive order (i.e. 1-A, 2-B, etc.). The test involves attentional skills, motor planning, and working memory [15]. This test is adopted for a wide range cases of brain dysfunction [61], moreover normative data are available for several countries according to relevant factors such as age, education level and sex [61]. Patients must complete the task as quickly as possible, if an error occurs then the examiner requests to correct it: this increases the total duration (time) thus reducing the final score assigned by the examiner (which is typically based only on the time spent). The test is asked to be performed without lifting the pen from the paper/tablet, however a wide set of in-air movement can be observed in the example reported in Figure 12 revealing the needing, for a mild dementia patient, to ideally retrace the path already written to be able to move forward from an error or hesitation point. The equivalence between the standard (pencil-paper) and the digital (pen-tablet) version of the TMT has been recently verified [76].

It is of interest to report that crossed pentagons, TMT and CDT tests have been recently adopted and compared within the context of handwriting processing to discriminate between HC, MCI, and AD [50]. To the aim the following features were considered: pressure, numbers of segments, velocity, acceleration, jerk, in-air and on-the-pad total time. Accuracy of pentagons, TMT-partA, TMT-partB abd CDT were, respectively, of 66.2%, 69.0%, 63.3% and 67.6%. The combination of all tasks was able to provide increased performance thus revealing they have a certain degree of complementarity.

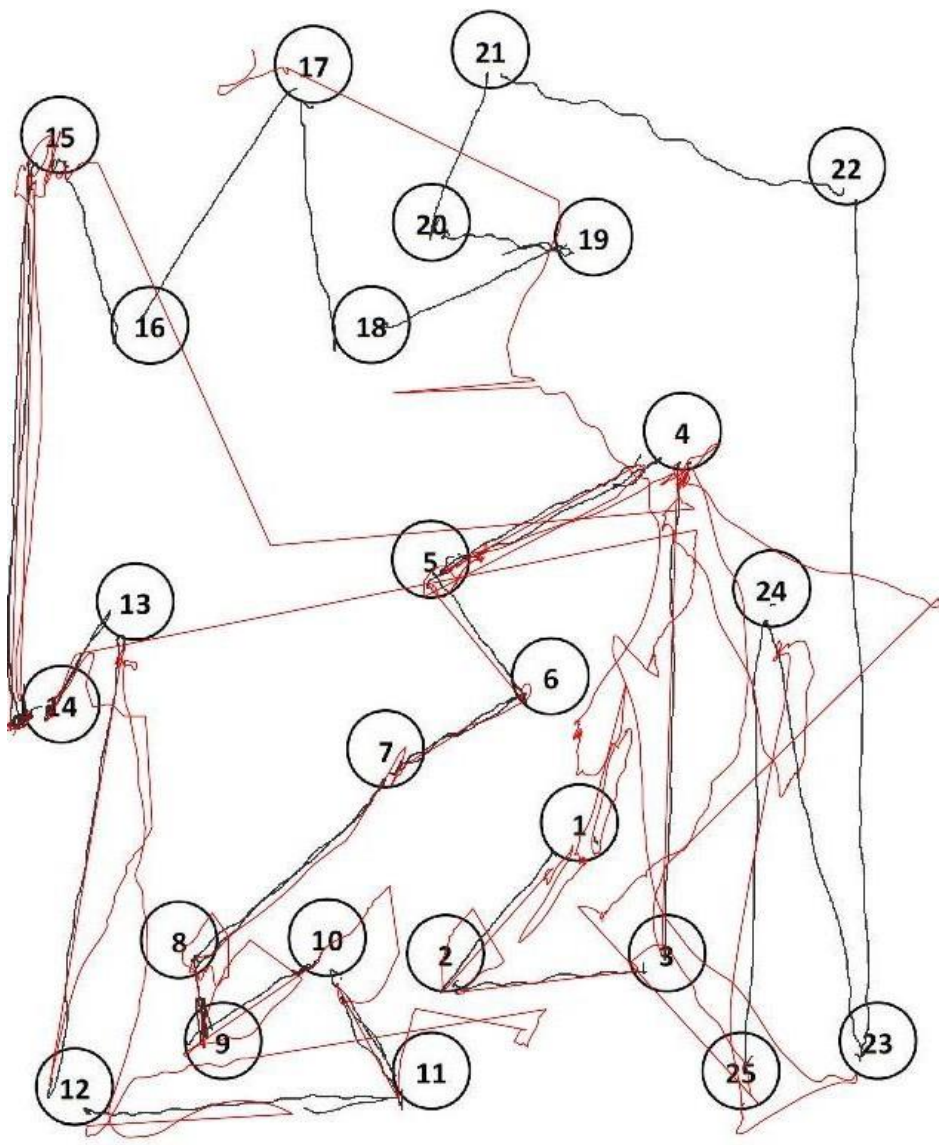

Fig. 12: Trail Making Test (the black color represents the on-surface movement, while the red color the in-air one)

*2.3.10 Cancellation test of digits*

These tasks are selective attention tests based on a cancellation task. The patient is asked to find targets (e.g. the number 5 in the example in Figure 13) within a short time constraint. So far, it has been reported that they are useful not only to discriminate AD and HC, but also to monitor the evolution of the cognitive decline [17]. Clinicians typically consider errors performed by patients; however, a digitized version of the test is also able to provide info related to the searching pattern (in air movement in Figure 13).

## 3 Keystroke/tactile/touch analysis

Keystroke dynamics have been extensively used for identification aims on physical keyboards [69, 5, 97] and recently on virtual keyboards when considering smartphones and tablets [98]. In this last situation a wider range of interactions can be considered including finger-swiping patterns drawing, touch-dynamics and, of course, signatures [47, 23]. It is evident that aspects involved in handwriting/signing described in the previous sections are not far from those involved in more general hand-based interaction tasks because they involve the same hand motor area within the brain [32]. So far it has been underlined that typewriting includes a cognitive phase, an associative phase and an autonomous phase [105].

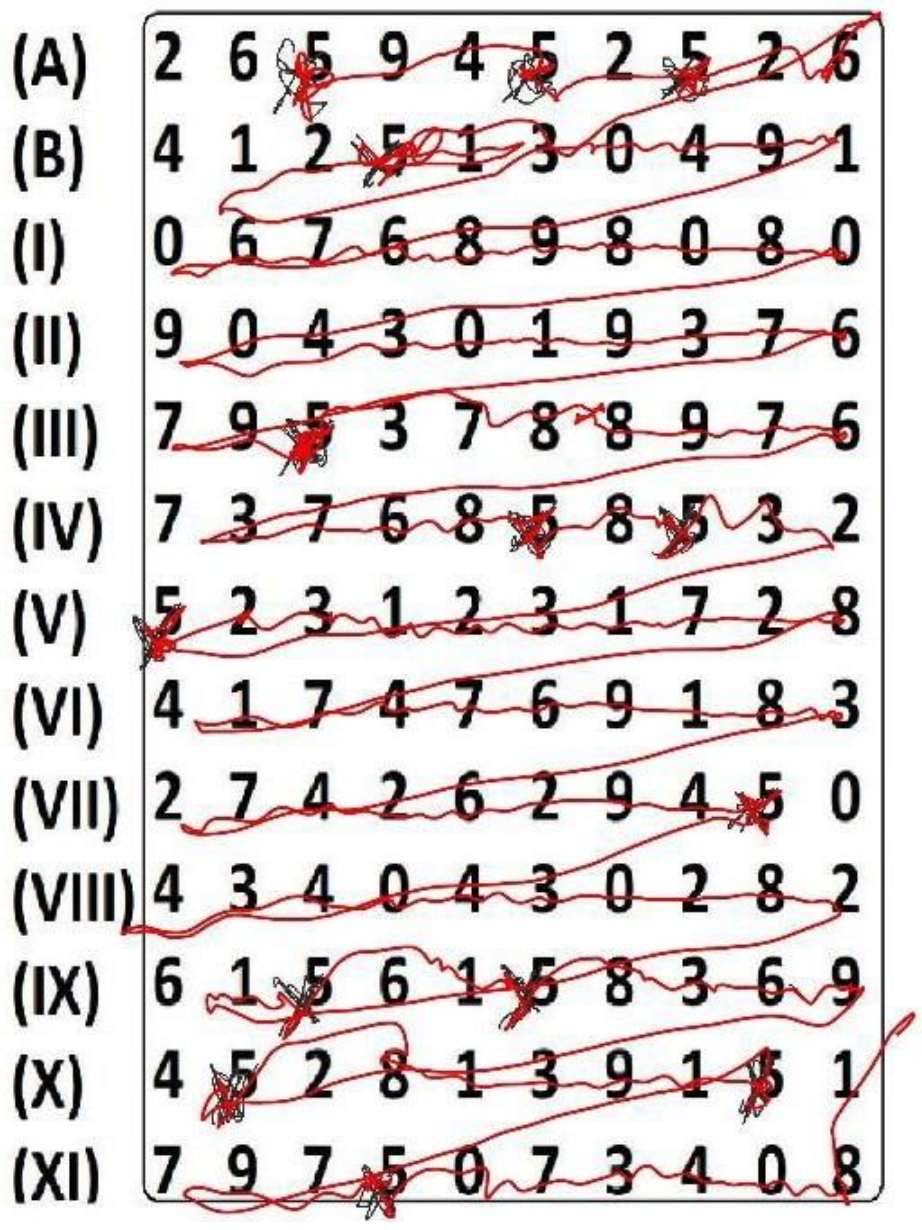

Fig. 13: Cancellation test of the digit '5' (the black color represents the on-surface movement, while the red color the in-air one)

Based on the previous observations, touch dynamics can be used also for health evaluation. However, it must be underlined that writing/drawing/signing and, more in general, interacting with a finger on a screen is different from using a pen/tablet system: dynamics as well as the final 2D drawing can be very different [47]. Main reasons are:

- habits in using the pen instead of the finger and/or vice versa;
- finger size compared to the pen's one (in terms of contact point);
- non-rigidity of the finger;
- friction between the finger and the screen.

It is worth noting that, so far, a useful tool for evaluating motor skills is finger-tapping, a test based on a special tool that allows to count the number of key taps within a given time interval (e.g. 30 seconds). This test is used for assessing the presence of bradykinesia, that is, an unnatural slowness in initiating and carrying out simple voluntary movements [93, 99]. Other interesting tasks can be considered. Iakovakis et al. [44] acquired fragmentary typing of short text on a touchscreen smartphone involving 18 PD patients and 15 HC. In this case features adopted were those of the typical key-stroke domain: hold times (time between the pressing and the releasing of a key), flight times (time between the releasing of a key and the pressing of the next one), etc. The adopted classification schema reported 0.82 and 0.81 of, respectively, sensitivity and specificity. Noyce et al. [74] adopted the following parameters for the PD vs HC classification: Kinesia Score (KS30) as, number of key taps in 30 seconds, Akinesia Time (AT30) as the mean dwell time on a key, Incoordination Score (IS30) as the variance of flight time between two consecutive keys and Dysmetria Scores (DS30) related to accuracy of key presses. It was observed that KS30, AT30 and IS30 were significantly able to discriminate PD patients from HC, moreover the same parameters were also correlated with UPDRS motor scores. Similar results have been obtained considering key hold time series and early PD patients [37]. Typing activity on smartphones, independently from the text, has been also considered [32]. In this last case, participants were requested to copy a randomly selected text for five minutes. The time sequence of flight times was used to compute parameter features to be fed to a set of different classifiers: a sensitivity/specificity of 0.81/0.81 has been reported in the binary PD/HC classification task. A very recent work has investigated and compared different touch gestures on the same device: flick, drag, handwriting, pinch, tap, and alternating finger tapping [100]. A wide set of spatial, velocity, time and pressure-based features was considered with the aim to distinguish between early PD patients and HC. The following results were observed: PD subjects resulted in less-efficient finger trajectories, less stable speed, less stable pressure and, higher tremor than HC. Touch gestures and typing appeared to be complementary tasks and an analysis of each task reported drag gestures the most performing one for classification aims. The best performance was achieved by using all categories of features. Lipsmeier et al. [58] considered also finger tapping (recording all touchscreen events) within a set of many tests related to the use of a smartphone (sustained phonation, rest tremor, postural tremor, balance and gait). The study involved 44 PD and 35 HC. The finger tapping appeared to be the less performing task, however it must be underlined that only intratap variability was considered as feature.

Problems in hand movements are often the first symptoms of neurological disorders, which do not include only PD, but also Essential Tremor (ET) and Huntington's disease (HD) [18, 92]. On the other hand, dementia diseases, as for example Alzheimer Disease (AD), first results in cognitive rather than motor degradation. In fact, it is well-known that complex tasks including cognitive load (e.g. clack drawing and pentagons) are generally considered [48]. However also coping tasks can be considered. Van Waes et al. [105] requested to a set of 20 young HC, 20 cognitively healthy elderly and 12 age-matched elderly with mild cognitive impairment (MCI) or mild dementia due to AD to perform a typing copy task. Different performances were observed among the three groups.

More in general, a comprehensive user analysis must involve the monitoring of multiple behavior including typing, menu navigation, swipes, drawing and activity understanding [95], [38]. Unfortunately, no works are still available in this direction considering a mobile device, so that it can be considered an open field of research. Very few works are available demonstrating the possibility of using touch dynamics for emotion recognition by considering common unlock Android touch patterns [4] or typing on touchscreen [66].

## 4 Tasks classification

According to our previous work [34], handwritten tasks can be classified into three categories:

- Mechanical tasks – with no cognitive effort, this task can be performed without any heavy load because it is a repetitive movement that the user can do in an automatic way. He is habituated to do it regularly in his life. This is the case of the handwritten signature, handwriting text in capital letters, and handwritten text in cursive letters. Usually, these kinds of tasks are quite straight forward and trivial. These kinds of tasks are quite frequent in the healthy population and find an important niche of applications in biometric recognition of people (user identification and verification [24]).
- Cognitive effort tasks – these tasks require some psychical effort to copy a complex drawing. In this case, he requires a strategy to start the task. Some cognitive aspects are important because he needs to know the parts of the drawing already done and the parts that are missing. This kind of task is especially challenging for those people affected by cognitive impairment, such as dementia, mild cognitive impairment, etc. This is the case of the house drawing task, pentagon drawing test, clock drawing test, Osterrieth complex figure test, trail making test, cancellation test of an specific digit and Bank-check copying. Clock drawing test is a special case, because no model is presented prior or during the task. It has to be imagined during performance.
- Fine motor control tasks – these tasks do not require a heavy cognitive load as the drawing itself is simple and straightforward to understand and memorize at a glance. However, good motor control is required to perform the task. This is the case of the Archimedes spiral drawing test, straight line test, spring drawing task, and concentric circle drawing test. Table 1 summarizes the best tasks for each application purpose. This table is focused on on-line acquired signals. Although the table presents security and health applications separately, it is important to point out that the same signal can reveal identity and pathologies. Thus, privacy is another interesting research topic that involves both, security and health [26].

Table 1: Summary of tasks classified by applications in the field of security and health

| Tasks | Security (user identification and verification) | Health |
|---|---|---|
| Signature | – classical application that is increasing popularity in on-line case (supermarkets, post office, etc.) [24, 2, 19, 47, 46]<br>– personality assessment [68]<br>– international competitions exist to compare different algorithms [43, 65, 112] | – although pathologies can be detected (e.g. Alzheimer's disease [79, 68, 82, 25, 110]) this is not a popular task in health applications requiring more investigation due to controversial results [8] |

| Tasks | Security (user identification and verification) | Health |
|---|---|---|
| Handwriting | – capital letters [89]<br>– cursive letters [88]<br>– letter level writer identification [12]<br>– gender recognition [91, 60, 59, 36, 1, 20]<br>– writer identification [111, 106, 107, 56]<br>– competitions on writer identification [20]<br>– competitions in gender identification [41] | – Parkinson's disease [71, 93, 21, 22, 77]<br>– Huntington's disease [7]<br>– developmental dysgraphia [33, 114]<br>– attention deficit hyperactivity disorder [54]<br>– autism spectrum disorder [84]<br>– obsessive-compulsive disorder [87]<br>– fatigue [34]<br>– depression, stress, etc. [57], although better results are found using drawing tasks<br>– drug abuse, such as Marijuana [29], alcohol [78], caffeine [102] |

| Tasks | Security (user identification and verification) | Health |
|---|---|---|
| Drawing | – graffiti's author identification (offline) [101,51]<br>– preliminary results in on-line case [62] | *Pentagon test*<br>– fatigue [34]<br>– Alzheimer's disease [35]<br>*Clock drawing test*<br>– Alzheimer's disease [96,52, 73,35]<br>– mild cognitive impairment [35]<br>– mild major depressive disorder [42]<br>*House drawing*<br>– Alzheimer's disease [35,72]<br>– mild cognitive impairment [35,72]<br>– hypoxemic patient analysis [28]<br>– fatigue [34]<br>*Archimedean spiral, meanders and straight lines*<br>– Parkinson's disease [86,85]<br>– Huntington's disease [7]<br>– essential tremor [63,80,70, 27]<br>– developmental dysgraphia [31]<br>– fatigue [34]<br>– brachial dystonia [81]<br>*Single or overlapped circles*<br>– Huntington's disease [7]<br>– schizophrenia [9]<br>*Spring task*<br>– fatigue [34]<br>– developmental dysgraphia [31,103]<br>– schizophrenia [9,14]<br>– bipolar disorder [14]<br>– Parkinson's disease [93]<br>– Huntington's disease [7]<br>*Rainbow task*<br>– developmental dysgraphia [31,67]<br>*Saw task*<br>– developmental dysgraphia [31] |

| Tasks | Security (user identification and verification) | Health |
|---|---|---|
| | | *Rey-Osterrieth complex figure test*<br>– mild cognitive impairment [11]<br>– Alzheimer's disease [53]<br>*Multiple geometrical figures copying*<br>– dementia [16]<br>*Trail making test*<br>– Alzheimer's disease [50]<br>*Cancellation test*<br>– Alzheimer's disease [17]<br>*Tree drawing*<br>– Alzheimer's disease [83]<br>– mild cognitive impairment [83] |

## 5 Conclusions

Handwriting is probably one of the most complex tasks that human beings can perform. In addition to being considered a personal behavioral trait (suitable for biometric recognition in security applications), it can also reveal health aspects (when analyzing its quality).

A large amount of scientific literature exist in both application fields: security and health. However, there is no unified activity to be performed by hand writers. Depending on the specific application field, there are some tasks that can unveil richer information than others. Thus, we have tried to systematically review the existing tasks and applications with the goal to serve as a guide for presenting the main alternatives and the topics where they have succeed.

In a first level, we can classify the tasks into three categories: signature, handwriting (cursive or capital letters), and drawings, being the latest one the richer in possibilities.

In a second level, we can classify the task into three different categories according to the specific skill required to perform the task: mechanical, cognitive effort and fine motor control. This second level of tasks is mainly relevant for health applications. However, in opposition to the first classification, this is not a disjoint set, as each task requires some amount of effort from the other classes. Thus, it just depicts the predominant effort.

While large amount of possible tasks exists, one research goal to be addressed is to find the best tasks for each application: those who require few realization time and provide good discrimination capability (for instance, to differentiate essential tremor from Parkinson disease).

We forecast new potential applications in the future based in online handwriting, especially in health. We encourage scientific community to test several handwriting tasks in order to find the optimal one. This paper summarizes the main successful ones and can serve as potential task catalogue to explore when studying new or existing problems.

**Acknowledgements** This work was supported by Spanish grant PID2020-113242RB-I00, and by grants NU20-04-00294 (Diagnostics of Lewy body diseases in prodromal stage based on multimodal data analysis), and PRIN2017 – BullyBuster project – A framework for bullying and cyberbullying action detection by computer vision and artificial intelligence methods and algorithms. CUP: H94I19000230006.

**Compliance with Ethical Standards** The authors declare that they have no conflict of interest.

All procedures performed in studies involving human participants were in accordance with the ethical standards of the institutional and/or national research committee and with the 1964 Helsinki declaration and its later amendments or comparable ethical standards. For this type of study formal consent is not required.

This chapter does not contain any studies with animals performed by any of the authors. Informed consent was obtained from all individual participants included in the study.